\documentclass[preprint,aps,floatfix,superscriptaddress,pre]{revtex4}

\usepackage{amsmath, amsthm, amssymb, graphicx}
\usepackage{color}
\usepackage{multirow}
\usepackage{ulem}
\usepackage{float}
\usepackage{bm}
\usepackage{subfiles}
\usepackage{siunitx}
\usepackage{booktabs}
\usepackage{array}

\begin{document}

\title{Critical phenomena in presence of symmetric absorbing states: a microscopic spin model with tunable parameters}
\author{Ahmadreza Azizi}
\affiliation{Department of Physics, Virginia Tech, Blacksburg, VA 24061-0435, USA}
\affiliation{Center for Soft Matter and Biological Physics, Virginia Tech, Blacksburg, VA 24061-0435, USA}
\author{Michel Pleimling}
\affiliation{Department of Physics, Virginia Tech, Blacksburg, VA 24061-0435, USA}
\affiliation{Center for Soft Matter and Biological Physics, Virginia Tech, Blacksburg, VA 24061-0435, USA}
\affiliation{Academy of Integrated Science, Virginia Tech, Blacksburg, VA 24061-0563, USA}
\date{\today}

\begin{abstract}
The Langevin description of systems with two symmetric absorbing states yields a phase diagram with three different
phases (disordered and active, ordered and active, absorbing) separated by critical lines belonging to
three different universality classes (generalized voter, Ising, and directed percolation). In this paper we present a 
microscopic spin model with two symmetric absorbing states that has the property that the model parameters can be varied in a continuous way.
Our results, obtained through extensive numerical simulations, indicate that all features of the Langevin description
are encountered for our two-dimensionsal microscopic spin model. Thus the Ising and direction percolation lines merge into a generalized
voter critical line at a point in parameter space that is not identical to the classical voter model.
A vast range of different quantities are used to determine the universality classes of the order-disorder and 
absorbing phase transitions. The investigation of time-dependent quantities at a critical point belonging to the
generalized voter universality class reveals a more complicated picture than previously discussed in the literature.
\end{abstract}

\maketitle

\section{Introduction}
Studies of phase transitions separating active from absorbing states have contributed
tremendously to our understanding of non-equilibrium physics \cite{Odo08,Hen08}. Many systems have a unique absorbing state
that can be accessed through an active-absorbing phase transition belonging to the directed percolation universality class,
but this is not the only possible scenario. 
Examples of systems with two symmetric absorbing states ($Z_2$ symmetry)
are found in numerous fields, ranging from opinion dynamics \cite{Vaz03,Cas09,Fer14} to linguistics \cite{Abr03,Bax09,Rus11}
and from catalysis \cite{Kra92,Fra96} to species competition \cite{Cli73,Shi19} and population ecology \cite{Bor12,Bor13}.
Often models discussed in this context are voter models where the current state of a discrete variable (which often is an Ising spin
that takes on the values $\pm 1$) depends on the distribution of states of other variables in a specified neighborhood.

A unique feature of the voter model is the absence of bulk noise which yields coarsening without surface tension \cite{Dor01,Tar15}.
The resulting slow dynamics is revealed by the fact that in two space dimensions the interface density $\rho$ decays logarithmically
with time, $\rho \sim 1/\ln t$ \cite{Sch88,Kra92,Fra96}. Remarkable features are also revealed at the active-absorbing phase transition
in systems with two symmetric absorbing states. Indeed, in critical $Z_2$ symmetric models without bulk noise both the interface
density and the magnetization (defined as the sum over all spins) display a logarithmic decay with time for the case that a symmetry-breaking
transition between an ordered and a disordered phase takes place at the same time as an absorbing phase transition
separating an active from an absorbing phase. Quantities in systems belonging to this
{\it generalized voter universality} class \cite{Ham05} are characterized by a set of critical exponents that differs from the 
corresponding set of the directed percolation universality class \cite{Odo08}.

Simulations of a kinetic Ising model in two dimensions provided early indications that a second scenario is possible for systems with two symmetric
absorbing states \cite{Dro03}: the splitting of the voter critical point into two separate transitions, namely an order-disorder transition
belonging to the universality class of the two-dimensional Ising model and an absorbing phase transition belonging to the directed 
percolation universality class. Al Hammal {\it et al.} \cite{Ham05} proposed a unifying picture of a phase diagram that contains three critical lines,
with the directed percolation and Ising lines merging into a generalized voter line. This was achieved through the investigation of the 
Langevin equation
\begin{equation}\label{eq_gvm}
\frac{\partial \phi}{\partial t}=D \nabla^2 \phi+(a\phi-b\phi^3)(1-\phi^2)+ \sigma \sqrt{1- \phi^2} \, \eta
\end{equation}
where $\phi$ is a space- and time-dependent coarse-grained field with continuous values between $-1$ and 1, whereas $\eta$ is a Gaussian
white noise with zero mean and unit variance. For $a=b=0$ this Langevin equation is identical to the Langevin equation for the
classical voter model \cite{Dic95,Mun97}. Numerical simulations of Eq. (\ref{eq_gvm}) reveal that the three critical lines
merge at a point that is not identical to the classical voter point $a=b=0$.

In numerical studies of kinetic Ising models \cite{Dro03,Vaz08,Azi18} the different scenarios predicted by the Langevin equation
are not found for a fixed range of the spin-spin interactions. Only when modifying the interaction range
can one observe a change in scenario, with a unique generalized voter phase transition for nearest neighbor
interactions and separate directed percolation and Ising phase transitions once the interaction distance exceeds a critical value.
Other systems were shown to allow access to some parts of the phase diagram obtained from the Langevin equation (\ref{eq_gvm}). 
Two-dimensional interacting monomer \cite{Par12} or monomer-dimer models \cite{Par15} were found to exhibit successive order-disorder and active-absorbing
phase transitions belonging to the Ising and directed percolation universality classes. In \cite{Ben16} a reaction-diffusion system was proposed 
that at the coarse-grained level is described by (\ref{eq_gvm}), but the physically meaningful
range of parameter values does not allow to access the different critical lines. In an interesting work \cite{Rod15}
Rodrigues {\it et al.} discussed a two-dimensional lattice model with freely tunable parameters. By construction their phase diagram contains a point where
their model is identical to the classical voter model. They recovered the same three phases as one obtains when simulating the Langevin equation (\ref{eq_gvm}).
However, in stark contrast to the Langevin description, the model of Rodrigues {\it et al.} only has two critical lines as a line of discontinuous phase transitions
is replacing the generalized voter critical line. In addition,
the Ising and directed percolation critical lines end exactly at the point where the classical voter model is realized, which is different from what is 
obtained from the Langevin equation.

In this paper we present a microscopic spin model with two continuously varying parameters that exhibits a phase diagram with all the phases and
all the critical lines obtained in the Langevin description. Our model is an extension of a model presented by Russell and Blythe  \cite{Rus11} that
displays a noise-induced dynamical transition between two coarsening regimes, one regime with curvature-driven coarsening (Ising-like 
coarsening) and one regime with coarsening in absence of surface tension (voter-like coarsening). We show that Eq. (\ref{eq_gvm}) provides
the correct stochastic equation of motion for our model. Through extensive numerical simulations we determine the universality classes
of the three different critical lines.

Our paper is organized in the following way. In Section II we present the microscopic spin model with tunable parameters and derive from
the Master equation the Langevin equation (\ref{eq_gvm}). In Section III we discuss the various quantities we studied in order to
determine the character of the different phase transitions and present data for each of the critical lines. We present a phase diagram
summarizing our results. Our study also reveals interesting transient properties at critical points belonging
to the generalized voter universality class. Finally, we conclude in Section IV.

\section{Microscopic model with tunable parameters}
In 2011 Russell and Blythe \cite{Rus11} presented a stochastic microscopic spin model that displays a transition between Ising-like (i.e. curvature
driven) coarsening and voter-like coarsening characterized by the absence of surface tension. They showed that the stochastic
equation of motion that follows from the microscopic model is given by the Langevin equation (\ref{eq_gvm}) with $b=0$:
\begin{equation}\label{eq_b0}
\frac{\partial \phi}{\partial t}=D \nabla^2 \phi+a\phi(1-\phi^2)+ \sigma \sqrt{1- \phi^2} \, \eta
\end{equation}
As we show in the 
following, this model can be extended in a way that the corresponding Langevin equation is given be Eq. (\ref{eq_gvm}) with $b$ being a 
tunable parameter that impacts the rate of going from one configuration to the next.

Russell and Blythe consider square systems composed of $L^2$ cells with $N$ spins in each cell. These classical spins can take on the values $\pm 1$.
A proposed update consists in replacing a randomly selected spin with a copy from a spin taken either from the same cell or from a neighboring cell.
The first move is related to the potential term in (\ref{eq_b0}), whereas the latter move is realizing diffusion.
The copy is taken from one randomly selected spin contained in one randomly selected neighboring cell with probability $q=\frac{h}{N}$ where $h$ is a system
parameter that turns out to be related to the diffusion constant $D$ in Eq. (\ref{eq_b0}). The probability that the copy is taken from an up respectively down spin
located in the same cell is $(1-q) p$ respectively $(1- q) (1-p)$ where $p \ne \frac{1}{2}$ introduces a systematic bias. In \cite{Rus11} the choice
$p = x_i + \frac{2 a}{N} x_i (1-x_i) (2 x_i -1) = \frac{1}{2} \left( 1 + \phi_i \right) + \frac{a}{2 N} \phi_i \left( 1 - \phi_i^2 \right)$ was made 
where $x_i = \frac{n_{i \uparrow}}{N}$ is the fraction of up spins in cell $i$ and $\phi_i = \frac{n_{i \uparrow}-n_{i \downarrow}}{N} = 2x_i -1$
is the local (cell) magnetization density. In addition, $n_{i \uparrow}$ respectively $n_{i \downarrow}$  is 
the total number of up respectively down spins in the cell and $a$ is a model parameter. 
It follows from the Fokker-Planck equation in the limit $N \longrightarrow \infty$ that (\ref{eq_b0}) is the correct stochastic equation of motion
for this model.

\begin{figure}
 \centering \includegraphics[width=0.45\columnwidth,clip=true]{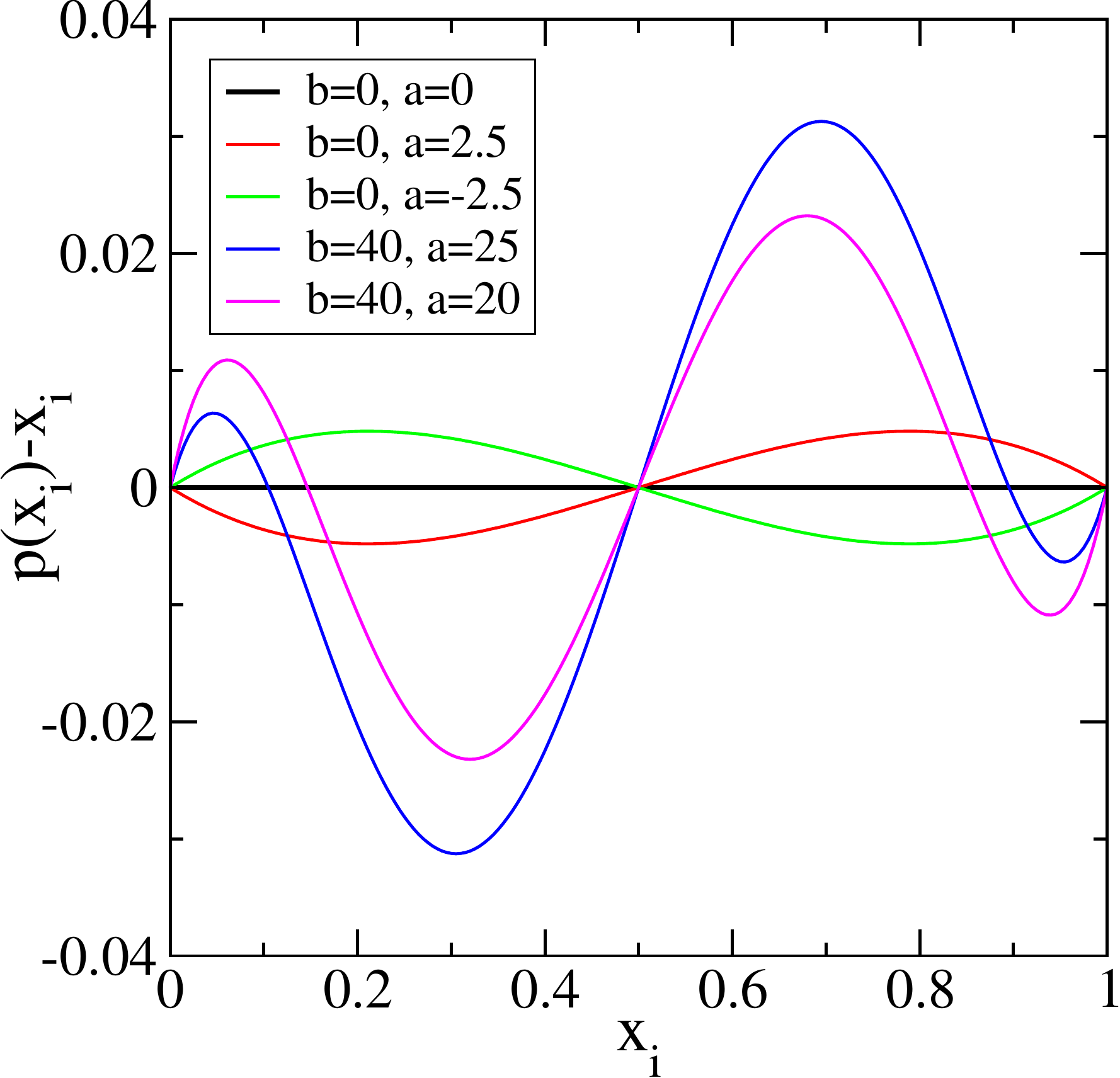}
 \caption{Deviations of the probability (\ref{eq_px}), to take as copy an up spin located in the same cell, from the straight line $y=x_i$ obtained
for the special case $a=b=0$. For $a$ or $b$ non-zero, $p(x_i)$ is no longer
simply proportional to the number of up spins in the cell, as for some values of $x_i$ the probability to select an up spin is enhanced, whereas for
other values of $x_i$ it is reduced.
}
\label{fig1}
\end{figure}

In order to obtain the full Langevin equation (\ref{eq_gvm}) in the limit $N \longrightarrow \infty$ the probability $p$ for taking the copy from 
the local cell has to be modified in an appropriate way. We propose to use the expression
\begin{equation} \label{eq_px}
p(x_i) = x_i + \frac{2 a}{N} x_i (1-x_i) (2 x_i -1) - \frac{2 b}{N} x_i (1-x_i) (2 x_i -1)^3
\end{equation} 
for this probability as a function of the fraction of up spins in the local cell, or, equivalently,
\begin{equation} \label{eq_pphi}
p(\phi_i) = \frac{1}{2} \left( 1 + \phi_i \right) + \frac{a}{2 N} \phi_i \left( 1 - \phi_i^2 \right) - \frac{b}{2 N} \phi_i^3 \left( 1 - \phi_i^2 \right)
\end{equation}
in terms of the local magnetization density $\phi_i$. For $a=b=0$, the probability to use for an update an up spin from the same cell is proportional to the fraction
of up spins in the cell, as $p(x_i) = x_i$. For $a$ or $b$ non-zero, we introduce a bias that for some values of $x_i$ enhances or reduces this probability when compared to
the case $a=b=0$. Fig. \ref{fig1} shows this bias when plotting $p(x_i)-x_i$ as a function of $x_i$ for some of the values of $a$ and $b$ used in this work.

The corresponding Langevin equation follows from the Fokker-Planck equation that results from the Kramers-Moyal expansion of the Master equation.
Noting that a move changes $x_i$ by the fixed value $\pm \Delta$ with $\Delta = \frac{1}{N}$, the transition probabilities entering the Master
equation are
\begin{eqnarray}
w\left( x_i + \Delta | x_i \right) & = & \left( 1 - x_i \right) \left\{ \frac{h}{4 N} \sum\limits_j x_j + \left( 1 - \frac{h}{N} \right) p(x_i) \right\} \\
w\left( x_i - \Delta | x_i \right) & = & x_i \left\{ \frac{h}{4 N} \sum\limits_j x_j + \left( 1 - \frac{h}{N} \right) p(1-x_i) \right\}
\end{eqnarray}
where $p(x_i)$ is given by Eq. (\ref{eq_px}), whereas the summation is over the four nearest neighbor cells $j$. With these transition probabilities,
a straightforward calculation yields in the limit $N \longrightarrow \infty$ the Fokker-Planck equation
\begin{equation} \label{eq_FP}
\frac{\partial P(x_i,t)}{\partial t} = - \frac{\partial}{\partial x_i} \left( v(x_i) P(x_i,t) \right) + \frac{1}{2} \frac{\partial^2}{\partial x_i^2}
\left( d(x_i) P(x_i,t) \right)
\end{equation}
with
\begin{equation}
v(x_i) = \frac{h}{4} \sum\limits_j \left( x_j - x_i \right) + 2 a x_i (1-x_i) (2 x_i -1) -2 b x_i (1-x_i) (2 x_i -1)^3
\end{equation}
and
\begin{equation}
d(x_i) = 2 x_i \left( 1 - x_i \right)~.
\end{equation}
Finally, using the substitution $\phi_i = 2 x_i -1$ in the Langevin correspondence yields the following stochastic
equation of motion:
\begin{eqnarray}
\frac{\partial \phi_i}{\partial t} & = &  a \phi_i \left( 1 - \phi_i^2 \right) - b \phi_i^3 \left( 1 - \phi_i^2 \right) + 
\frac{h}{4} \sum\limits_j \left( \phi_j - \phi_i \right) + \sqrt{2} \sqrt{ 1 - \phi_i^2} \, \eta
\end{eqnarray}
which is identical to the Langevin equation (\ref{eq_gvm}) after noting that the sum over $j$ is proportional to the discretized version of the diffusion
term $\nabla^2 \phi$. We note that the parameters $a$ and $b$ showing up in the probabilities (\ref{eq_px}) and (\ref{eq_pphi}) are in fact determining the
shape of the local potential in the Langevin equation.

\section{Numerical phase diagram and universality classes}

In the following we show that the Russell-Blythe model with the probability (\ref{eq_pphi}) indeed exhibits 
a phase diagram with a disordered state, an absorbing state and an ordered (but not absorbing) state, separated by phase transition
lines belonging to the generalized voter, the two-dimensional Ising and the directed percolation universality classes.
Most of the data discussed in the following have been obtained for $L=100$ and $N=100$ (i.e. with a total of one million spins) 
and periodic boundary conditions in both directions.
We fix $h=0.5$ and vary the system parameters $a$ and $b$ in order to obtain the phase diagram in the $b-a$ plane.
Our time step corresponds to $L \times L \times N$ updates, i.e. on average every spin is updated once 
during a time step.

\subsection{Quantities}

As we expect to encounter transition lines that either separate an active from an absorbing phase or a disordered from an ordered phase,
we need to use very different quantities in order to probe the universality classes of the different transitions.

Useful quantities for locating an active-absorbing phase transition point and determining the corresponding critical universality class \cite{Odo08}
are provided by the time-dependent average number of flipped cells $N_f(t)$ and the survival probability $P_s(t)$ of flipped cells, where
a flipped cell is a cell where at least one of the spins is changed when compared to the initial state.
Indeed, exactly at the absorbing phase transition these two quantities are expected to vary algebraically with time,
\begin{equation}
P_s(t) \sim t^{-\delta} ~~~\mbox{and}~~~ N_f(t) \sim t^\eta~,
\end{equation}
with the directed percolation values $\delta_{DP} = 0.45$ and $\eta_{DP} = 0.23$ \cite{Gra96,Voi97} and the 
generalized voter universality class values
$\delta_{GV}=1.0$ and $\eta_{GV}=0$ \cite{Dic95,Odo04}. Starting 
point for our calculation of these quantities is an initial state where all spins in all the cells take on the same value (say, $-1$,
which results in the local (cell) magnetization density $\phi=-1$). Having prepared the system in this way,
we flip all the spins in four neighboring cells to $+1$, yielding the cell magnetization density
$\phi = 1$ for these cells, and then update the system using the scheme described in Section II. At every time step we measure
the number of cells with $\phi \ne -1$. Repeating this procedure thousands of times with different random number sequences, the
average number $N_f$ of flipped cells (i.e. cells with $\phi \ne -1$) 
then results from an average over these different runs. For the survival probability $P_s$
we record at each time step $t$ the fraction of runs that still have cells with $\phi \ne -1$.

At the voter critical point the interface density $\rho$ displays a characteristic time dependence in the 
form of a logarithmic decay \cite{Dor01}:
\begin{equation}
\rho(t) \sim 1/ \ln t~
\end{equation}
which makes it an excellent quantity for identifying the location of the critical point as well as the universality class.
We prepare our system in a fully disordered initial state (i.e. every spin can take on the values 1 and $-1$ with the same probability 1/2)
and then calculate the interface density
\begin{equation}
\rho(t) = 1 - \langle \frac{1}{2L^2} \sum\limits_{j=1}^L \sum\limits_{k=1}^L \phi_{j,k}(t) \left( \phi_{j+1,k}(t) + \phi_{j,k+1}(t) \right) \rangle
\end{equation}
where $\langle \cdots \rangle$ indicates an ensemble average over different noise realizations and $\phi_{j,k}(t)$ is the cell magnetization 
density at time $t$ of the cell $i = \left(j, k \right)$ where $j$ and $k$ are labeling the cells along the horizontal and vertical
direction respectively.

Another dynamic quantity of interest is the two-time autocorrelation function
\begin{equation}
C(t,s) = \langle \frac{1}{L^2} \sum\limits_{j=1}^L \sum\limits_{k=1}^L \phi_{j,k}(t) \phi_{j,k}(s) \rangle -
\langle \frac{1}{L^2} \sum\limits_{j=1}^L \sum\limits_{k=1}^L \phi_{j,k}(t) \rangle \, 
\langle \frac{1}{L^2} \sum\limits_{j=1}^L \sum\limits_{k=1}^L \phi_{j,k}(s) \rangle
\end{equation}
which at a critical point is expected to display the simple aging scaling \cite{Hen10}
\begin{equation} \label{eq_aging}
C(t,s) = s^{-\tilde{b}} f_C(t/s)
\end{equation}
when starting from a disordered initial state.
The scaling function $f_C$ only depends on the ratio of the two times $t$ and $s$ and decays algebraically for large 
arguments with an exponent $\lambda$. In \cite{Azi18} we showed that
this aging scaling indeed prevails at a voter critical point, with the exponents taking on the values $\tilde{b}= 0.120(5)$ and
$\lambda = 1.00(2)$. 

Finally, we also computed standard quantities used for investigating equilibrium second order phase transitions, notably
moments of the magnetization density $M$ and the fourth order cumulant (Binder cumulant) of the magnetization \cite{Bin81a,Bin81b,Sel06}
\begin{equation} \label{eq_binder}
U=1 - \frac{\overline{M^4}}{3 \overline{M^2}^{\,2}}~.
\end{equation}
As we will see below, the Binder cumulant is very useful for finding as a function of the parameter $a$ the location of the paramagnetic-ferromagnetic 
phase transition for fixed value of $b$. In order to obtain these equilibrium quantities, we first let the system relax for several thousand time steps
before sampling data at fixed time intervals. This is repeated using different random number sequences so that $\overline{\cdots}$ indicates
both a time and an ensemble average.

\subsection{Phase diagram of the microscopic spin model}
The phase diagram that results from the Langevin equation (\ref{eq_gvm}) for systems with two absorbing states is well understood \cite{Ham05,Vaz08}.
There exists a special value $b = b_s$ that separates the phase diagram into two parts with very different properties.
For $b < b_s$ a unique phase transition takes place at a critical value of the parameter $a$. This phase transition, which separates a disordered, active 
steady state from an ordered and absorbing state, belongs to the universality class of the generalized voter model. For $b > b_s$, however, two
consecutive phase transitions are encountered when increasing $a$. The system first undergoes an order-disorder phase transition that belongs to the
universality class of the two-dimensional Ising model. The two phases separated by this phase transition are active phases. Increasing $a$ further, a second
phase transition to an absorbing state takes place. This second transition belongs to the directed percolation universality class.

\begin{figure}
 \centering \includegraphics[width=0.45\columnwidth,clip=true]{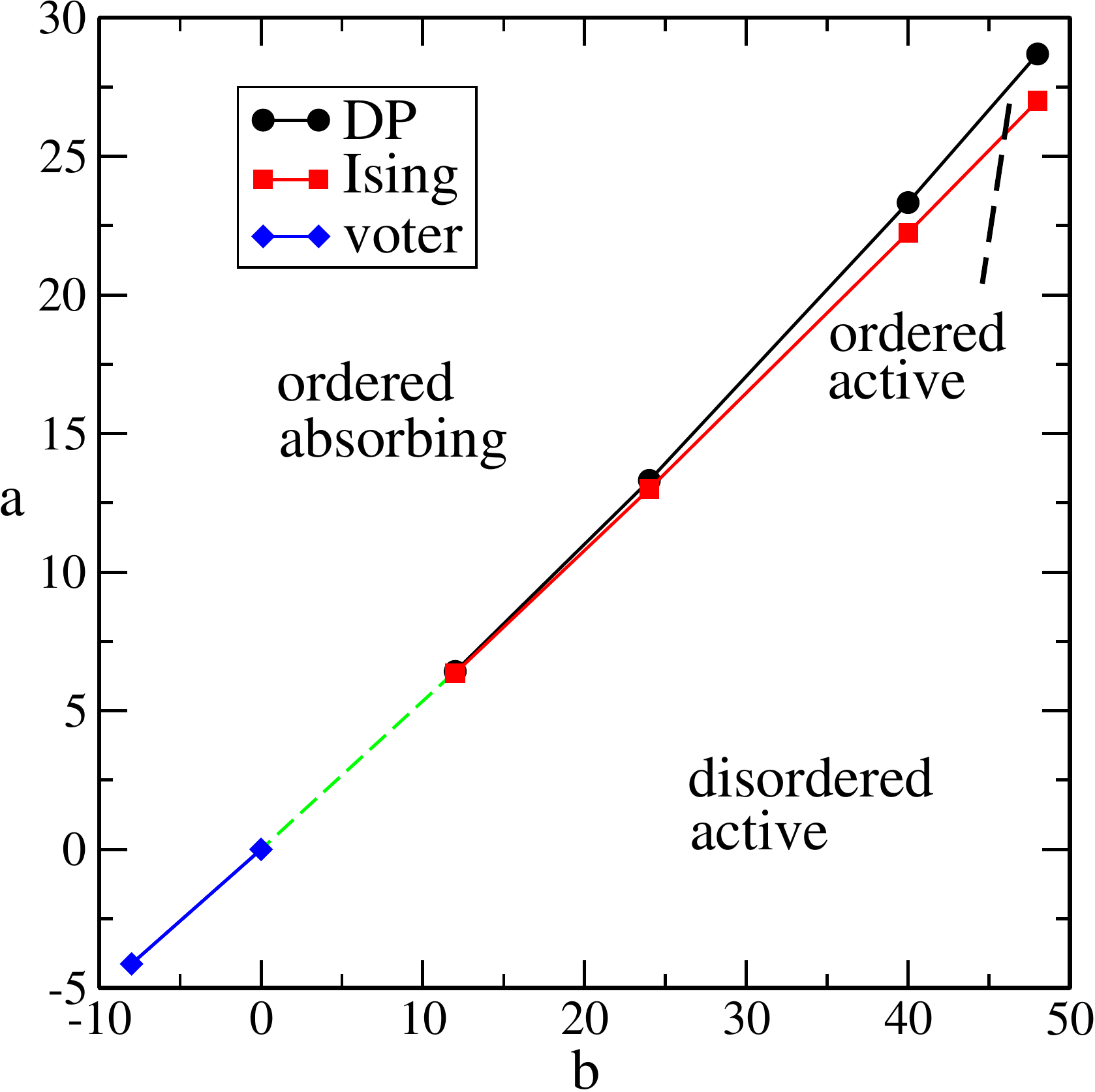}
 \caption{Phase diagram in the $b-a$ plane obtained form simulations of the microscopic spin model.
The properties of the phase transitions taking place at particular values of the system parameters
$a$ and $b$ (filled symbols) have been elucidated. Along the green dashed segment a special point exists where all three
critical lines merge. No attempt was made to locate this special point. The full 
lines are guides to the eye.}
\label{fig2}
\end{figure}

Previous efforts to verify these scenarios through microscopic spin models have only been partially successful, due to the lack of a microscopic spin model with
parameters that can be changed continuously. Indeed, the best studied cases are that of absorbing Ising models on a square 
lattice with the added voter rule that a spin is not allowed 
to flip if it points in the same direction as all the spins in a specified neighborhood \cite{Dro03,Vaz08,Azi18}. For nearest neighbor interactions only a voter
transition is observed, whereas extending the interaction range to further neighbors yields at some point a splitting of the voter critical point and the
appearance of two phase transitions. However, for a system of spins on a lattice the interaction range can only be changed in discrete steps (going from 
nearest neighbor interactions to next nearest neighbor interactions etc.) which does not make these models convenient ones for a more systematic study.

For the model presented in the previous Section the parameters $a$ and $b$, whose values can take on any real number, directly enter into the probabilities to update
the system. As a result changing the values of these two parameters allows to fully explore the phase diagram and fine tune the properties of the model.
We show in Fig. \ref{fig2} the phase diagram that we have obtained from the numerical simulations of this microscopic spin model. We encounter three different
steady states: a disordered and active phase in the lower part of the phase diagram, an ordered and inactive phase in the upper part of the diagram, and an intermediate
phase that is both ordered and active and that for values of $b$ larger than a special value $b_s$ is intercalated between the other two phases. As we show in the rest
of this Section, the critical lines separating these phases belong to the expected three universality classes (generalized voter, two-dimensional Ising, and directed percolation)
in agreement with the results obtained from the Langevin equation. These three lines should meet at a special point $(b_s, a_s)$ located in the Figure
along the green dashed line. We did not attempt to locate precisely this special point through a time-consuming systematic
search. We stress that for our model the different phases are readily encountered with nearest-neighbor interactions only, whereas in the model discussed in \cite{Vaz08}
the interaction range has to be fine tuned as the splitting of the voter critical point is only observed when the interaction range is extended to 
at least third nearest neighbors.
 
\subsection{$b \le 0$: one critical line belonging to the voter universality class}

We start our analysis of the character of the phase transitions by looking at the case $b \le 0$. For these values of $b$ we encounter a single phase
transition that is expected to belong to the generalized voter universality class.

\begin{figure}
 \centering \includegraphics[width=0.45\columnwidth,clip=true]{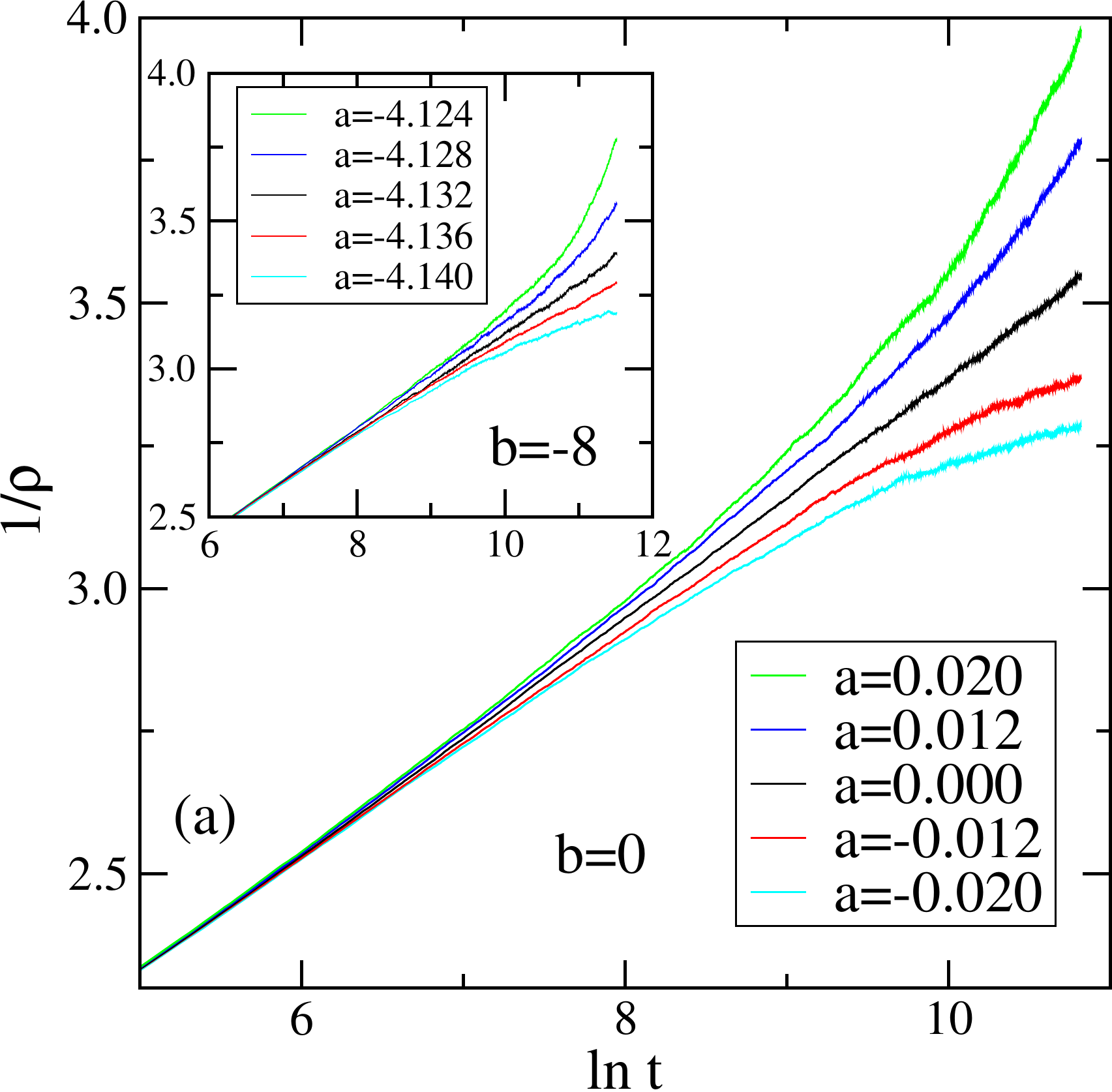}\hspace*{0.3cm}
 \includegraphics[width=0.45\columnwidth,clip=true]{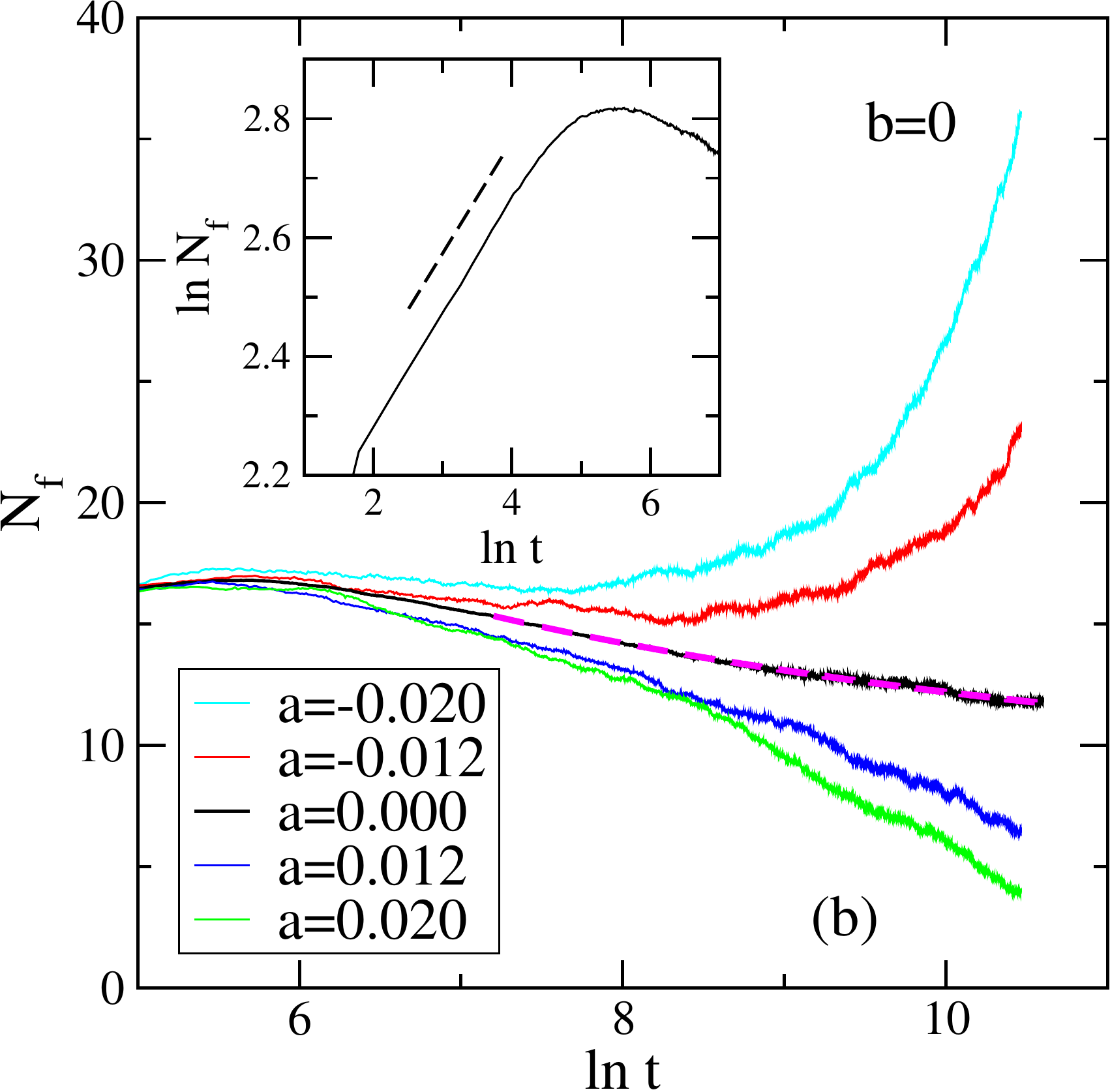}\\[0.2cm]
 \centering \includegraphics[width=0.45\columnwidth,clip=true]{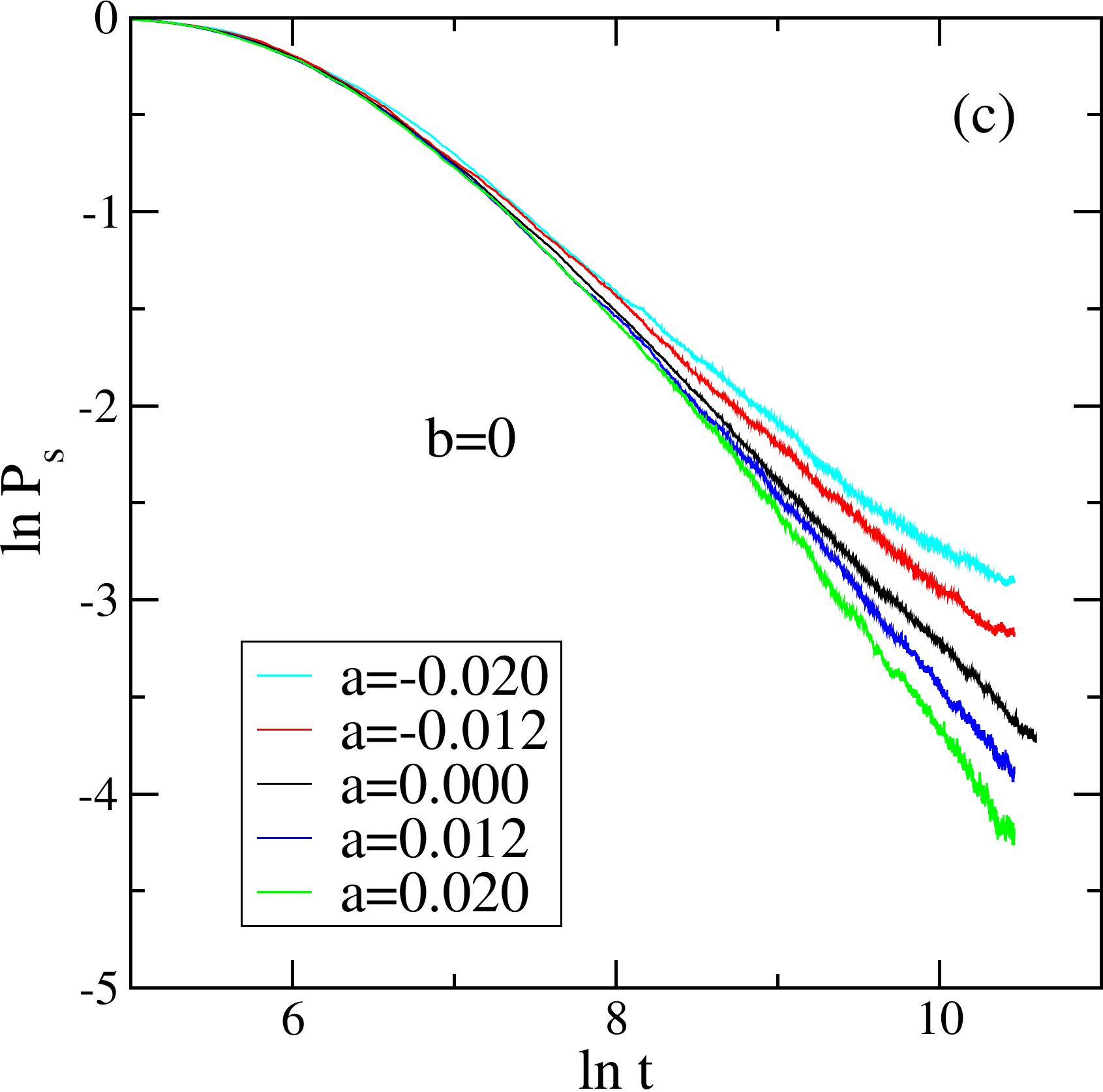}
 \caption{Time dependence of various quantities for systems with $b=0$ and different values of $a$.
Based on these data a phase transition belonging to the generalized voter universality class takes place at $a=0$.
The different quantities display a typical behavior:
(a) the interface density $\rho$ decays as $1/\ln t$, (b) the average number of flipped cells $N_f$ also decays 
like $1/\ln t$ (the best fit $80.74/\ln t$ is shown as the magenta dashed line on top of the $a=0$ line), 
which is compatible with $\eta_{GV}=0$, and (c) the survival probability decays algebraically 
with an exponent $\delta_{GV} \approx 1$. The same 
behavior is encountered for $b < 0$. This is illustrated in the inset of panel (a) for the case $b = -8$
for which the phase transition is found to take place at $a=-4.132(2)$.
The inset in panel (b) shows that at early times the average number of flipped cells increases at the critical point $a=0$ algebraically 
with an effective exponent 0.19 (indicated by the dashed line)
The data for the interface density result from averaging over at least 3000 independent runs. For the
other two quantities averages were done over at least 5000 independent runs.}
\label{fig3}
\end{figure}

In order to locate the critical value of $a$ for a fixed value of $b$ we investigate how different quantities behave when changing $a$. This is
shown in Fig. \ref{fig3} for the case $b=0$. The inset in Fig. \ref{fig3}a shows that the same behavior is encountered for other values of $b \le 0$.
The interface density $\rho$ in Fig. \ref{fig3}a and the survival probability $P_s$ in Fig. \ref{fig3}c directly allow to locate the phase transition 
and confirm the voter critical behavior through their characteristic time dependence ($1/\rho(t) \sim \ln t$ and $P_s(t) \sim t^{-\delta_{GV}}$ with $\delta_{GV}
\approx 1$).

An interesting behavior is exhibited by the average number of flipped cells $N_f$ in Fig. \ref{fig3}b. The expected voter critical behavior for this quantity
($N_f(t) \sim t^{\eta_{GV}}$ with $\eta_{GV} = 0$) suggests that $N_f(t)$ should be constant, and this has indeed been observed in a two-dimensional absorbing Ising
model \cite{Vaz08} as well as in linear voter models \cite{Cas12,Azi18}. 
For our model, however, $N_f$ shows an early algebraic growth before it decreases as $1/\ln t$ at late times. While this is formally compatible with $\eta =0$, this 
behavior indicates that different scenarios are possible at a generalized voter critical point. As the number of cells whose cell magnetization density 
is not equal to $-1$ can be viewed as a proxy for the magnetization of a system formed by $L^2$ cells, we note that the logarithmic time dependence
in Fig \ref{fig3}b is mirrored by a corresponding logarithmic behavior of the magnetization of non-linear models at their voter critical point
\cite{Dor01,Cas12,Azi18}. In the inset of Fig. \ref{fig3}b we show the interesting early time behavior of $N_f$ at the critical point. The observed
power-law increase with an effective exponent 0.19, indicated by the dashed line,
is reminiscent of the initial slip behavior of the magnetization in systems with a small, non-vanishing magnetization
quenched to an equilibrium critical point \cite{Jan89,Jan92}. After this initial regime, $N_f$ reaches a maximum and then decays logarithmically with time,
as discussed above. 

It is worth pointing out that this algebraic increase is not explained by the known properties of the generalized voter universality class.
Indeed, the well-known hyperscaling relation $\eta_{GV} = \frac{d z_{GV}}{2}-2 \delta_{GV}$ \cite{Dic95,Mun97} 
only allows to recover the value $\eta_{GV} = 0$, known to be correct for critical linear voter models,
when plugging in the values $\delta_{GV} =1$ and $z_{GV} = 2$ for the survival probability exponent and the dynamic exponent. 
The observation of an early-time increase of the average number of flipped cells,
together with the previously discovered non-constant magnetization at the critical points of non-linear models, indicates that our theoretical
understanding of critical non-linear voter models is incomplete.

\begin{figure}
 \centering \includegraphics[width=0.45\columnwidth,clip=true]{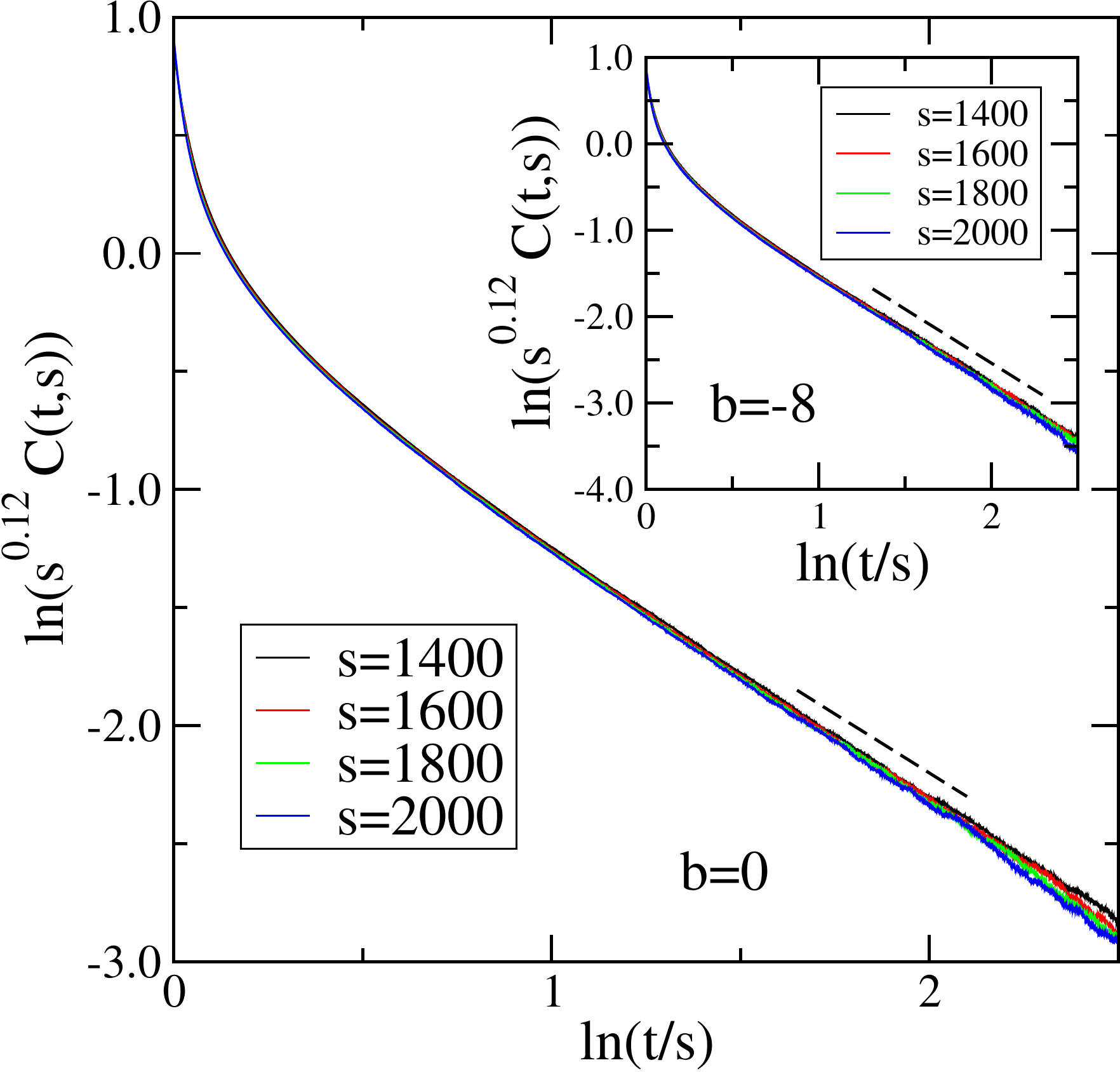}
 \caption{Scaling of the autocorrelation function $C(t,s)$ for $a=0$ and $b=0$. Simple aging scaling (\ref{eq_aging}) with the exponents $\tilde{b}=0.12$ 
and $\lambda = 1$ (indicated by the dashed line) is encountered, in agreement with the expected behavior at a voter critical point \cite{Azi18}. The inset shows that the same
behavior prevails for $b=-8$ and $a=-4.132$. These data result from averaging over 5000 independent runs.}
\label{fig4}
\end{figure}

In Fig. \ref{fig4} we verify that at the points $(b=0, a=0)$ and $(b=-8, a=-4.132)$ we have for the two-time
autocorrelation function the same critical aging scaling behavior
(\ref{eq_aging}) with exponents $\tilde{b} = 0.12$ and $\lambda =-1$ as encountered at other voter critical points \cite{Azi18}.
Taken together, the data shown in Figs. \ref{fig3} and \ref{fig4} unambiguously reveal that for $b \le 0$ our microscopic spin model
exhibits a line of generalized voter critical points separating the disordered active phase from the absorbing phase.

\subsection{$b > b_s$: two critical lines belonging to the Ising and directed percolation universality classes}

For fixed $b > b_s$ the expected scenario is the appearance of two separate phase transitions when varying $a$. In the following we
discuss the case $b=40$, but similar results are obtained for other values of $b$, see Fig. \ref{fig2}.

\begin{figure}
 \centering \includegraphics[width=0.45\columnwidth,clip=true]{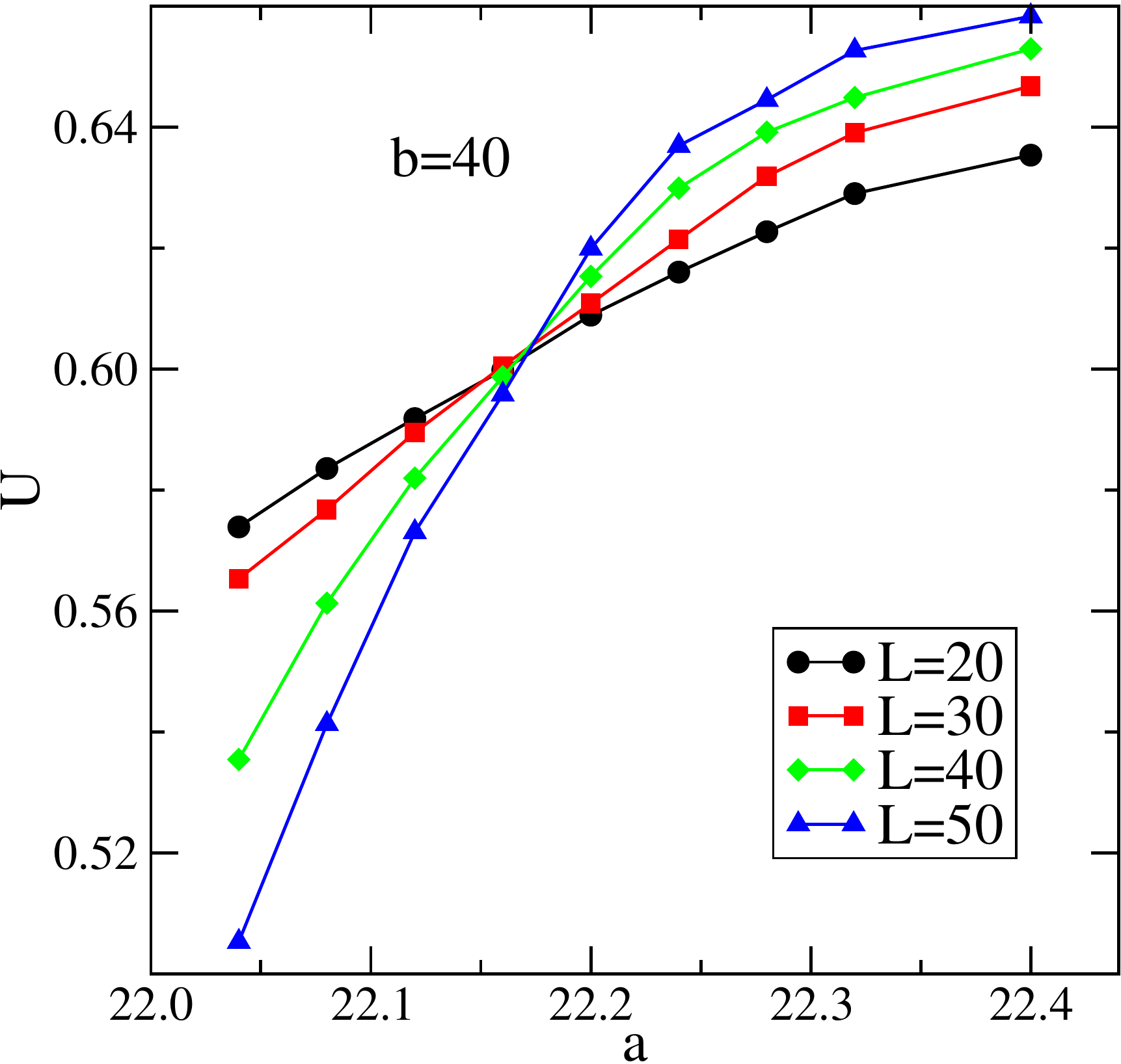}
 \caption{Binder cumulant $U$ for various numbers of cells $L^2$ as a function of $a$ when $b=40$. The different lines
intersect close to $a= 22.16$. The value of $U$ at this point is close to the universal value $U_I \approx 0.61$ for the
two-dimensional Ising universality class \cite{Kam93,Sel06}. For these equilibrium simulations the first $300,000$ time steps
were discarded before sampling data for a time average. The final data result from an additional averaging over 
typically 1000 different runs.
Error bars are comparable to the symbol sizes. The lines are guides to the eye. 
}
\label{fig5}
\end{figure}

In Fig. \ref{fig5} we show the Binder cumulant (\ref{eq_binder}) as a function of $a$ for various numbers of cells $L^2$. The dependence on the
system parameters and the system size is as expected for a continuous order-disorder phase transition. The lines for the different system sizes
cross in the vicinity of $a=22.16$ which provides an estimate for the location of the phase transition. The value of $U$ at that point is close to the
theoretical value $U_I \approx 0.61$ for a phase transition belonging to the universality class of the two-dimensional Ising model. The Ising
character of this phase transition is also verified through the behavior of the magnetization close to this point (not shown).

\begin{figure}
 \centering \includegraphics[width=0.45\columnwidth,clip=true]{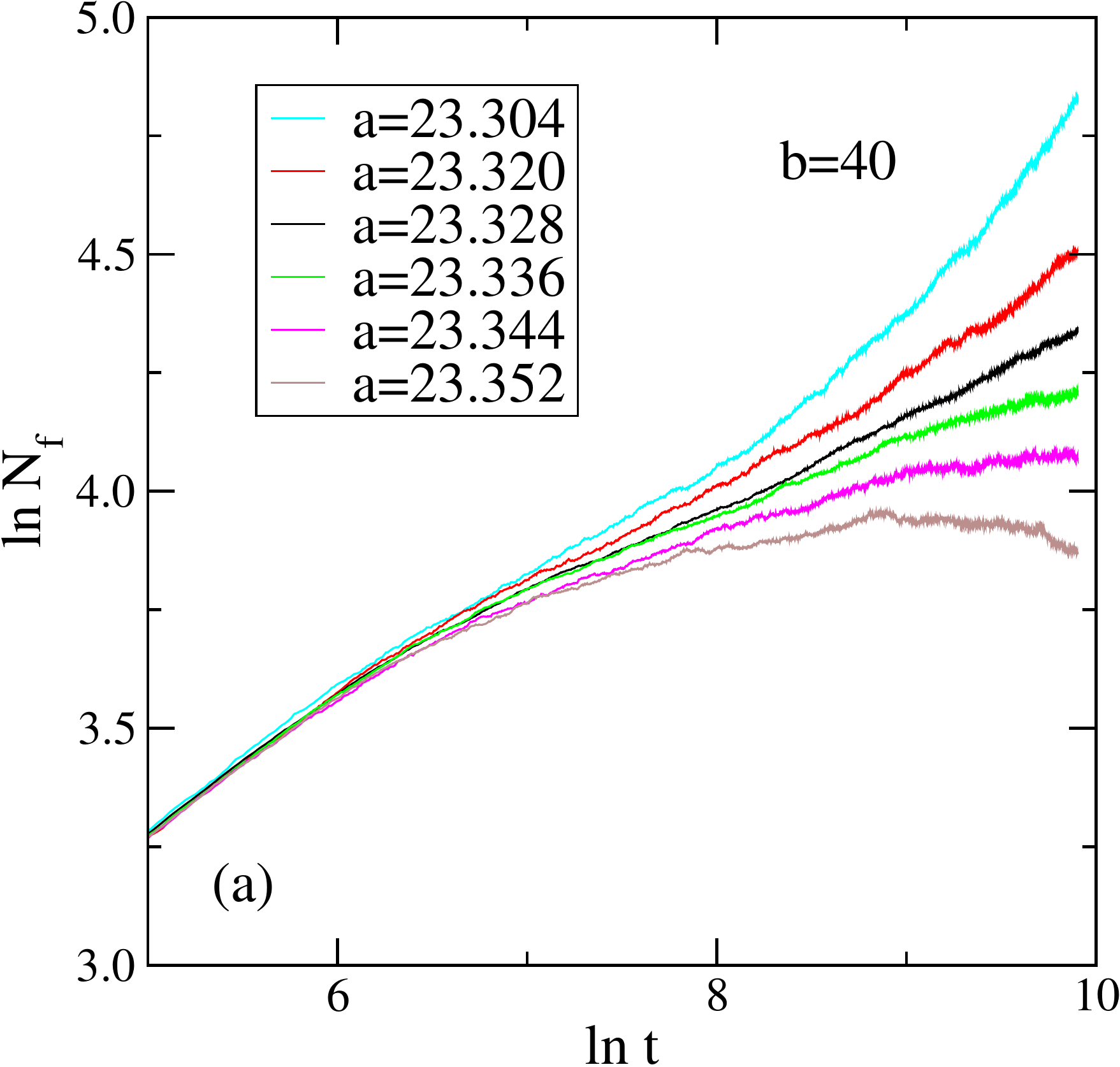}\hspace*{0.3cm}
 \includegraphics[width=0.45\columnwidth,clip=true]{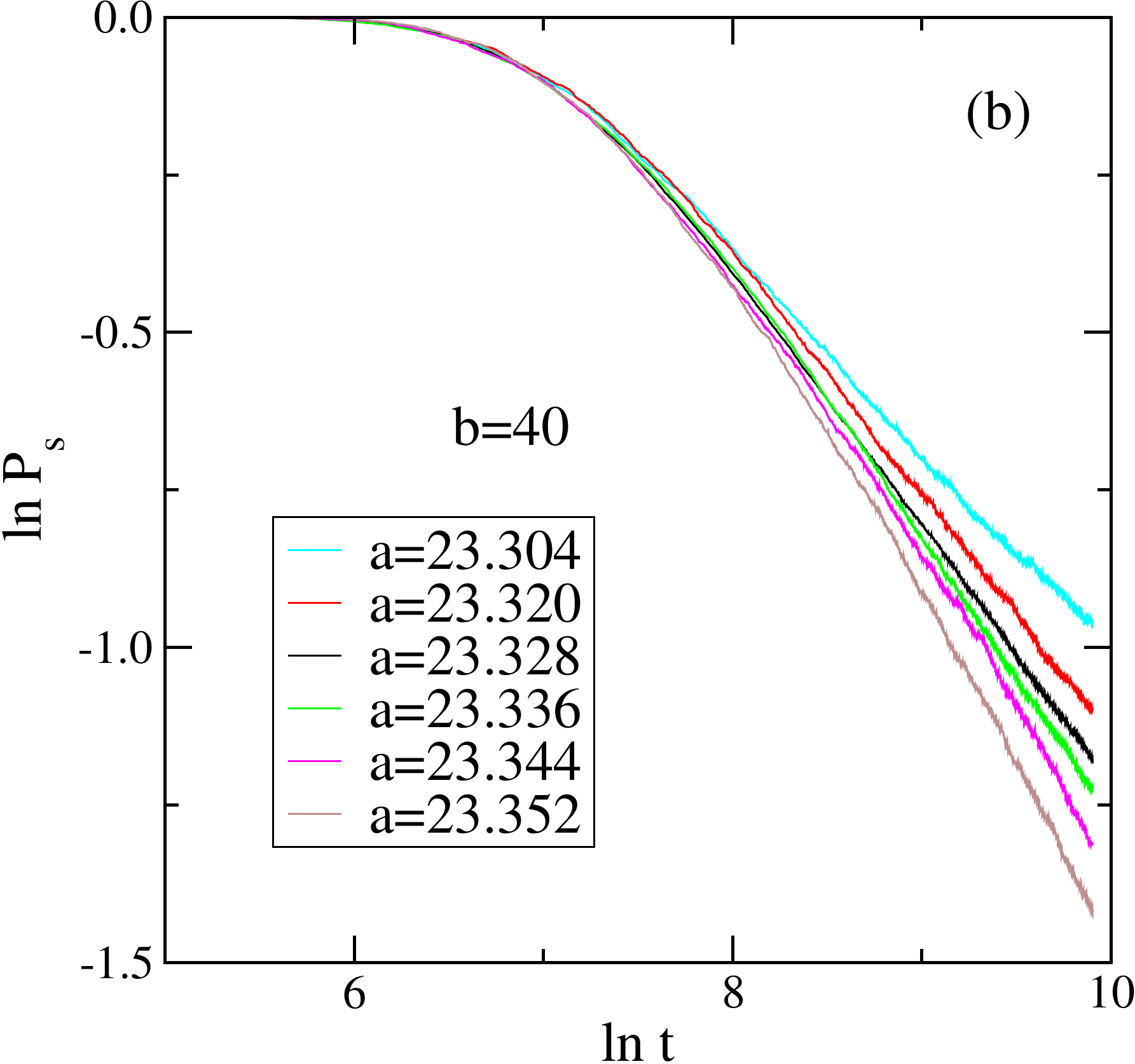}
 \caption{Time dependence of (a) the average number of flipped cells $N_f$ and (b) the survival probability $P_s$
for $b=40$ and different values of $a$. These data indicate that a directed percolation phase transition takes place
for $a =23.332(4)$. The data result from averaging over at least 4000 different realizations of the noise.
}
\label{fig6}
\end{figure}

Increasing further $a$ results in a second phase transition, this time between an ordered but active phase, the intermediate phase shown in
Fig. \ref{fig2}, and the absorbing phase. Both $N_f$ and $P_s$ display a power-law dependence on time around $a=23.332$,
see Fig. \ref{fig6}. The corresponding 
exponents ($\eta = 0.20 \pm 0.02$ for $N_f$ and $\delta = 0.43 \pm 0.03$ for $P_s$) are very close to the expected values for directed percolation
($\eta_{DP}=0.23$ and $\delta_{DP} = 0.45$).

\section{Conclusion}
Systems with two symmetric absorbing states have found applications in fields ranging from population ecology to opinion dynamics. 
The presence of symmetric absorbing states allows for two different scenarios when crossing from the active phase to the absorbing
phase. In a direct transition, where the order-disorder transition coincides with the active-absorbing transition, critical properties
are those of the generalized voter universality class, characterized by the fact that after a quench to the critical point
the interface density decays logarithmically with time. The transition to the absorbing state can
also happen in two steps, with first an order-disorder transition belonging to the Ising universality class followed by a an absorbing
transition belonging to the directed percolation universality class. In the Langevin description a unified picture was provided
where the Ising and directed percolation critical lines merge into the generalized voter critical line \cite{Ham05}.

In this paper we discussed a microscopic spin lattice model with continuously tunable parameters that allows to
access all three critical lines predicted by the Langevin equation (\ref{eq_gvm}) without the need to fine tune 
the interaction range. Investigating various quantities used commonly in studies
of equilibrium and non-equilibrium critical phenomena, we clearly identified the different universality classes. A generalization to a larger
number of absorbing states (mirroring the extension from kinetic Ising to kinetic Potts models \cite{Dro03,Azi18}) is possible.

Our study revealed an early-time increase of the number of flipped cells (which can serve as a proxy for the
magnetization density) when quenching the system to a voter critical point, which came as a surprise to us. 
Indeed, from the known properties of the generalized voter universality class, the number of flipped cells
is expected to be time independent, a property indeed recovered at the critical points of linear voter models.
The situation is more complex for critical non-linear models, with $N_f$ displaying an effective initial power-law increase,
followed by a decrease with a logarithmic dependence on time. This is consistent with the earlier observation of
non-constant magnetization in critical non-linear voter models \cite{Dor01,Cas12,Azi18}. All this indicates
that our understanding of the critical properties of non-linear voter models is incomplete and warrants a
more in-depth investigation in the future.

In our current investigation we did not attempt to locate precisely the point at which the three critical lines merge. With a
microscopic spin model with parameters that can be varied continuously now at hand, it is only a question of available resources whether this point can be located
with precision high enough that allows an in-depth study of the critical properties at this special point. We plan to come back to this
intriguing issue.

\begin{acknowledgments}
Research was sponsored by the US Army Research Office and was accomplished under
Grant Number W911NF-17-1-0156. The views and conclusions contained in this document
are those of the authors and should not be interpreted as representing the official policies,
either expressed or implied, of the Army Research Office or the US Government.
\end{acknowledgments}

\end{document}